\documentclass[twocolumn,preprintnumbers,amsmath,amssymb,superscriptaddress]{revtex4}
\usepackage{graphicx}
\usepackage{dcolumn}
\usepackage{bm}
\usepackage{natbib}
\usepackage{multirow}
\usepackage{array}

\def\ltsima{$\; \buildrel < \over \sim \;$}
\def\simlt{\lower.5ex\hbox{\ltsima}}
\def\gtsima{$\; \buildrel > \over \sim \;$}
\def\simgt{\lower.5ex\hbox{\gtsima}}

\def\kms{{\,\rm kms^{-1}}}
\def\hompc{{\,h\,\rm Mpc^{-1}}}
\def\mpcoh{{\,h^{-1}\,\rm Mpc}}

\def\gpcubed{\,h^{-3}\mathrm{Gpc}^{3}}

\def\[{\begin{equation}}
\def\]{\end{equation}}
\def\m@th{\mathsurround=0pt }
\def\eqalign#1{\null\,\vcenter{\openup1\jot \m@th
 \ialign{\strut\hfil$\displaystyle{##}$&$\displaystyle{{}##}$\hfil
 \crcr#1\crcr}}\,}

\begin{document}
\title{Difficulties Distinguishing Dark Energy from Modified Gravity via Redshift Distortions}

\date{\today}
\newcommand{\ud}{\mathrm{d}}
\newcommand{\fpe}{f_\perp}
\newcommand{\fpa}{f_\parallel}
\newcommand{\om}{\Omega_m}

\author{Fergus Simpson}
 \email{frgs@roe.ac.uk}
\author{John A. Peacock}
\affiliation{SUPA, Institute for Astronomy, University of
Edinburgh, Royal Observatory, Blackford Hill, Edinburgh EH9 3HJ}

\date{\today}

\begin{abstract}

The bulk motion of galaxies induced by the growth of cosmic structure
offers a rare opportunity to test the validity of general relativity
across cosmological scales.  However, modified gravity can
be degenerate in its effect with the unknown
values of cosmological parameters. More seriously, even the
`observed' value of the RSD (redshift-space distortions) used to measure the
fluctuation growth rate depends on the assumed cosmological parameters
(the Alcock-Paczynski effect).  We give a full analysis of
these issues, showing how to combine RSD with BAO (baryon acoustic oscillations)
and CMB (Cosmic Microwave Background) data, in order
to obtain joint constraints on deviations
from general relativity and on the equation of state of dark energy
whilst allowing for factors such as non-zero curvature.  In particular we note that the evolution of $\Omega_m(z)$, along with the Alcock-Paczynski effect,
produces a degeneracy between the equation of state $w$ and the
modified growth parameter $\gamma$. Typically, the total marginalized
error on either of these parameters will be larger by a factor $\simeq 2$
compared to the conditional error where one or other is held fixed.
We argue that future missions should be judged by their Figure of Merit as defined in the $w_p - \gamma$ plane, and note that the inclusion of spatial curvature can degrade this value by an order of magnitude.

\end{abstract}
\maketitle

\section{Introduction}

Models of dark energy leave a characteristic signature embedded in
both the cosmic expansion and structure formation histories. Recent
observational progress has been made with the former, due to its
relative ease of measurement, leading to a measurement of the dark
energy equation of state parameter $w\equiv P/\rho c^2$ with better
than 10\% precision \cite{2008arXiv0803.0547K,2009arXiv0907.1660P}.
This work is geometrical, and so probes dark energy only through its
influence on the evolving expansion rate of the Universe. It is thus
possible that dark energy may be an illusion, indicating the need to
revise general relativity and thus also the Friedmann equation.  In
either case, the phenomenological dark energy term may well differ
from a cosmological constant ($w=-1$), and may change its equation of
state with redshift. These possible degrees of freedom need to be
allowed for before we can claim any evidence for a deviation from
general relativity. This paper thus considers how we can make
simultaneous measurements of the properties of dark energy and of
modified gravity.

A number of probes are capable of measuring $w$ via its influence
on the redshift-distance relation. This measurement alone is
effectively completely degenerate with a modification of gravity
on the scale of the Hubble radius. But for many models, the
Mpc scales of galaxy clustering
may be affected in a different way; the growth rate of
density fluctuations has thus emerged as a key means
of breaking this degeneracy between gravity and dark energy
\cite{2005PhRvD..72d3529L}\cite{2008Natur.451..541G}.
It is rather more difficult to
study the growth rate, due to uncertainty in the behaviour of galaxy
bias, but there are currently two promising avenues available for future
exploration. Weak gravitational lensing provides a direct measurement
of the dark matter distribution, and its evolution with redshift. It can also probe broader aspects of modified gravity, particularly the balance between perturbations to the time and space parts of the metric \cite{beanism,2008PhRvD..78f3503J}.
The focus of the present work will be the alternative technique, known as redshift-space distortions (RSD), which exploit the relationship between the large-scale coherent velocities of galaxies and the growth rate of perturbations.  

In real space, we expect the clustering of galaxies to be
statistically isotropic. However, in redshift space the
line-of-sight component of a galaxy's peculiar velocity breaks
this symmetry. Inside a virialized cluster of galaxies, the
orbital velocity dispersion scatters galaxy redshifts, creating
the `Fingers of God', and thereby erasing spatial information on
small scales. Across larger scales, galaxies coherently fall out
of voids and into overdense regions, considerably amplifying the
power in redshift space. These two effects are often treated
independently, although a more complex model is required to attain
a higher degree of precision \cite{scocciz}. For the present
purpose, the large-scale effect is the aspect of interest, since
continuity relates coherent peculiar velocities directly to the
growth rate of density fluctuations.

Observations to date have led to estimates of the growth rate at
various redshifts, although not yet at a useful level of precision
\cite{2002MNRAS.332..311H,2005MNRAS.361..879D,2007MNRAS.381..573R,2008Natur.451..541G}. Future
surveys are likely to cover orders of magnitude larger volumes,
thereby delivering the precision needed to discriminate
interesting models of modified gravity. But we shall see that, when
approaching this target, it may no longer be appropriate to make the
simplifying assumptions adopted to date.

In \S\ref{sec:sig} we review the process of determining the growth
rate from redshift distortions, before constructing a Fisher
matrix. Our results are presented in \S\ref{sec:results}, while in
\S\ref{sec:merit} we consider the implications for the proposed dark
energy Figure of Merit.


\section{Signatures of modified gravity} \label{sec:sig}

For any theory of gravity that could play the role of dark energy,
there is little reason to believe the formation of structure on large
scales would match that of General Relativity. A simple
phenomenological model to quantify such a deviation has been suggested
by Linder \cite{2005PhRvD..72d3529L} (see also Wang \& Steinhardt \cite{1998ApJ...508..483W}), parameterising the growth of
linear density perturbations as 
\[
f(z)\equiv\frac{\ud\ln\delta}{\ud\ln a}\simeq\Omega_{m}^{\gamma}(z),
\]
where $\delta$ is the fractional density fluctuation and $a$ is the
scale factor; this approximation typically holds to a precision of
$\sim0.1\%$. This explicitly restricts the value of $f(z)$ to unity at high
redshift, but this is unlikely to be problematic given that dark
energy only appears to be of cosmological significance at low
redshift. An exception to this would require a rather contrived
functional form of $w(z)$, one that maintains a significant amount of
dark energy at high redshift yet not sufficient to modify our
observation of the CMB. Observational limits on these `early dark
energy' models have been studied in \cite{earlyde}.

In the context of $\Lambda$CDM, $\gamma$ takes the value of
0.55. Galaxy surveys are sensitive to this parameter via the Kaiser
effect \cite{1987MNRAS.227....1K}, which in its simplest form is given
by
\[
\label{eq:pkkaiser}
P(k_{\parallel},k_{\perp})=P(k)\left(1+\beta\mu^{2}\right)^{2},
\]
where $\mu=k_{\parallel}/|\textbf{k}|$, the parameter $\beta(z)$ is defined as
\[
\beta(z)=\frac{f(z)}{b(z)}  ,
\]
and where we must restrict ourselves to the
large-scale linear regime of scale-independent
bias, or be prepared to model non-linear redshift distortions.
Given that our application of ( \ref{eq:pkkaiser}) will extend beyond General Relativity, it is important to ensure that the validity of this formalism remains intact. In arriving at the above equations we have implicitly assumed a continuity relation linking the velocity and density fields. This relies upon the comservation of comoving matter, and as such should hold under all metric theories of gravity.

Whilst the linear bias $b(z)$ is not directly observable, it can
be inferred either from the bispectrum or from the amplitude of galaxy
clustering -- where on large scales $\xi_{gg} = b^2 \xi_{mm}$ and for a given cosmology
$\xi_{mm}(z)$ is known from the CMB. Any claimed deviation from
$\gamma = 0.55$ will inevitably be met with great scepticism unless the
methodology is highly robust, and the bispectrum is unlikely to match
this requirement, given that it is inherently a nonlinear
quantity. We thus consider the argument from clustering amplitude:
$b=\sigma_{\rm gal}(z)/\sigma_8(z)$, where
$\sigma_{\rm gal}$ denotes the observed fractional rms in
galaxy number density. A measurement of
$\beta$ from redshift-space distortions thus yields
$f(z) \sigma_8(z)$ if we assume that
$\sigma_{\rm gal}$ can be measured with negligible uncertainty.

The matter fluctuation $\sigma_8(z)$ is of course not an
observable, but it can be inferred for a given choice of
cosmological parameters by taking the CMB as a reference point.
This effectively allows us to deduce $f(z)$ in terms of
observed fluctuations and known growth laws:
\[
f(z)=\Omega_{m}^{\gamma}(z)=\beta(z)\,\frac{\sigma_{\rm gal}(z)}{\sigma_{8}(z_{\rm CMB})}\,
\frac{G(z_{\rm CMB})}{G(z)} ,
\label{eq:getbias}
\]
where the universal linear growth function is $\delta\propto G(z)$.
This argument neglects the weak dependence of the last-scattering redshift
on the cosmological parameters; a more precise version is given below, following
equation (\ref{eq:fullpk}).

\subsection{Alcock-Paczynski}

Provided the distance-redshift relation is well known, the values of
$f \sigma_8(z)$ and $\sigma_{8}(z_{\rm CMB})$ may be extracted from redshift
distortions and the CMB respectively.  Note however that this
assumes perfect knowledge of the
background expansion history, $H(z)$, allowing us to map the true
observables (angles and redshifts) onto $k$~space.  Strictly speaking,
the $f\sigma_{8}$ term is thus \emph{not} directly observable; the actual
measurement inevitably incorporates corrections from the
Alcock-Paczynski effect \cite{1979Natur.281..358A}.

Similarly, the form of the power spectrum $P(k)$ presented in
$(\ref{eq:pkkaiser})$, commonly used in the analysis of redshift
distortions \cite{2001Natur.410..169P}\cite{2008Natur.451..541G},
implicitly assumes knowledge of the distance-redshift relation. In
reality the difference in the true functions $D_{A}(z), H(z)$ may
differ from our adopted fiducial values $\hat{D}_{A}(z)$,
$\hat{H}(z)$, leading to the two scaling factors
\[
 \label{eq:fp1}
\fpe=D_{A}(z)/\hat{D}_{A}(z)  ,
\]
\[ \label{eq:fp2}
\fpa=\hat{H}(z)/H(z)  ,
\]
which generate apparent wavenumbers $k'_\perp = \fpe k_\perp$ and
$k'_\parallel = \fpa k_\parallel$, where the prime denotes the coordinate system derived from the assumed cosmology. Thus the assumed cosmology changes the inferred value of $\beta$ \cite{1996MNRAS.282..877B}, and will also alter the value of $\sigma_{\rm gal}$ deduced from the data.

Guzzo et al. \cite{2008Natur.451..541G} propose an iterative method to
converge on the correct cosmology. While this proves useful for
independently determining $\Omega_m$ alone, it is unlikely to succeed
when extending the parameter set to include $w_0$,
its evolution $w_a\equiv - dw/da$, and the
global curvature $\Omega_k$.

\subsection{The apparent power spectrum}

Our chosen parameter set $p_i$ consists of
\[
[w_0, w_a, \Omega_\Lambda,  \Omega_k, \Omega_m h^2, \Omega_b h^2, n_s, A_s, \beta, \gamma,
  \sigma_p].
\]
The main cosmological parameters are taken to have fiducial values as
derived from WMAP5 \cite{2008arXiv0803.0547K}. An assumption about the redshift dependence of
bias is required, and we take $b(z)=0.6(1+z)$. Combined with $\gamma =
0.55$, this determines the value of $\beta$ for a given redshift.
Finally, $\sigma_p$ denotes the one-dimensional rms pairwise
velocity dispersion, which we take to be $300\kms$ --
or $3\mpcoh$ when converted to length units.

Why include both $w$ and $\gamma$ as free parameters? A more limited parameter set may lead us to misinterpret a simple dark energy fluid as a sign of modified gravity: quantities such as $[w_0, w_a, \Omega_k]$ all have an impact upon our inferred value of $\gamma$ by modifying the function $\Omega_m(z)$. If $\gamma$ does indeed deviate from $0.55$, there is no longer any reason to expect that the dynamical evolution of the universe precisely mimics that of a cosmological constant. It is therefore important to consider variation in the effective equation of state $w(z)$, in other words, the model that reproduces the evolution of the cosmic expansion. A recent claimed detection of modified gravity suffers from precisely this unjustified assumption of $w=-1$  \cite{beanism}.

The size of the parameter space can be reduced by one if we focus
on $w_p$, the value of $w$ at the pivot redshift. In the usual
linear evolution model, this is $w(a)=w_p + w_a(a_p-a)$, where
the pivot era $a_p$ is chosen so that errors in $w_p$ and $w_a$
are uncorrelated. In principle $w_a$ is an important parameter,
since detection of $w_a\neq0$ would disprove the cosmological
constant hypothesis. But in practice it is rather poorly measured,
and the present analysis is not greatly changed if we marginalize
over it. 

The isotropic real-space matter power spectrum, $P(k)$, is generated
following the HALOFIT \cite{halofit} prescription, and we extend (\ref{eq:pkkaiser})
to incorporate a simple model of the nonlinear redshift distortions.
\[
P(k_{\parallel},k_{\perp})= P(k)\left(1+\beta\mu^{2}\right)^{2}D(k\mu\sigma_{p}).
\]
where D is a Lorentzian given by
\[
D(k\mu\sigma_{p})=\frac{1}{1+\left(k\mu\sigma_{p}\right)^{2}/2}
\]
Owing to deviations between the assumed cosmology and the true
cosmology, our longitudinal and tangential coordinates are rescaled by
factors of $\fpa$ and $\fpe$ respectively, as given by (\ref{eq:fp1})
and (\ref{eq:fp2}). The apparent power spectrum $P^{\prime}(k')$ is
recast in the form below, as outlined in (A8) from Ballinger et
al. \cite{1996MNRAS.282..877B} (see also Matsubara \& Suto \cite{1996ApJ...470L...1M}):

%
\[ \label{eq:fullpk}
\eqalign{
P'_{\rm gal}(k')&=\frac{1}{\fpe^{2}\fpa}b{}^2P_{m}\left(\frac{k'}{\fpe}\sqrt{1+\mu'^{2}\left(\frac{1}{F^{2}}-1\right)}\;\right) \cr
&\times \left[1+\mu'^{2}\left(\frac{1}{F^{2}}-1\right)\right]^{-2} \cr
&\times \left[1+\mu'^{2}\left(\frac{\beta+1}{F^{2}}-1\right)\right]^{2}D\left(\frac{k'_\parallel \sigma_{p}}{\fpa}\right),
}
\]
%
where $\mu=k_{\parallel}/|\textbf{k}|$ and $F\equiv\fpa/\fpe$.  

We emphasise that the bias parameter here is the `true' bias, and we
do not need to define an `apparent' value: the rationale for this equation is that the amplitude of apparent galaxy number density fluctuations is unchanged by the Alcock-Paczynski transformation, and only the direction of wavevectors is altered.

Seo \& Eisentein \cite{2003ApJ...598..720S} present an equivalent calculation to the above, which involves evaluating the distortions first before applying the transformation, thereby simplifying the form to 

\[
P'_{\rm{gal}}(k')=\frac{1}{\fpe^{2}\fpa} b{}^2 P_{m}(k) \left(1+\beta\mu^{2}\right)^{2}D(k\mu\sigma_{p}) .
\]

This should give rise to equivalent results, but we prefer the
approach of (\ref{eq:fullpk}) in which the Alcock-Paczynski corrections
are exhibited explicitly. This has several advantages:
it makes it clear that there is a potentially strong degeneracy
between $\beta$ and $F$, as discussed by Ballinger et al. \cite{1996MNRAS.282..877B},
and it allows us to show directly the impact of the
geometrical corrections on the RSD signal, as discussed below.
 
Our model for the fiducial galaxy bias is simply parameterised as 
$b(z)=0.6(1+z)$. Within a given redshift bin, it is assumed that there is
negligible scale or redshift evolution, although in reality we may
typically expect a change of $\sim 10\%$ across a bin of width $0.2$.
We do not treat the bias as a nuisance parameter to be marginalized over,
since its perturbed value is given exactly for a given set of parameters:
\[
b = \frac{\Omega_m^{\gamma}(z)}{\beta}  .
\]
Adjustment of these parameters would thus in effect also change
the amplitude of the apparent power spectrum $P'_{\rm gal}(k')$.
We have experimented with an alternative parameter set, in which
$\beta$ is replaced by $b$, and find that our overall results are unchanged,
as required.

This simple parameterisations of the redshift
distortions, $\beta$ and $\sigma_{p}$, should suffice for this initial
exploration of statistical uncertainties. In practice, of course, there
would be the concern that the model may  prove too simplistic to apply in
detail, leading to a systematic bias in the results. This is an
important issue, but one that does not need to be explored here.

\subsection{Constructing likelihood contours}

As usual, we predict parameter uncertainties using the
Fisher-matrix formalism. The Fisher matrix (expectation of the
Hessian matrix of 2nd derivatives of $\ln{\cal L}$) is constructed
by numerical integration of the following expression
\cite{2009MNRAS.tmp..925W}, up to a cut-off at $k_{\rm{c}} = 0.3
\hompc$.
\[
F_{ij}=\frac{1}{4 \pi^2}\int_0^{k_c} \!\!\!
\int_0^{k_c} \left(\frac{\partial\ln P'}{\partial p_{i}}\right)\left(\frac{\partial\ln P'}
{\partial p_{j}}\right)V_{\rm{eff}}(\vec{k}) k'_{\perp}\ud k'_{\perp}\ud k'_{\parallel}.
\]
Here, the effective volume of the survey compensates for the shot
noise, as defined by \cite{Tegmark1998}:
\[
V_{\rm{eff}}\equiv V_{0}\left(\frac{\bar{n}P}{1+\bar{n}P}\right)^{2}.
\]
We emphasise that this is an integration over the full apparent
power spectrum defined above; in this way, the
Alcock-Paczynski effects are fully included.

The galaxy power spectrum alone will not yield well-defined
cosmological conclusions.
In order to include constraints from the CMB, we add the Planck
Fisher matrix defined by the DETF\footnote{http://www.physics.ucdavis.edu/DETFast/DETFast.zip}.
As usual in such work, the DETF Fisher matrix uses a
different parameter set from our preferred choice, and
so the matrix has to be subject to a coordinate transformation,
using the Jacobian matrix between one parameter set and
another. The final Fisher matrices used in this analysis
can be found at {\tt www.roe.ac.uk/{\tt\char'176}frgs/wgamma.html}.

\begin{figure*}
\includegraphics[width=80mm]{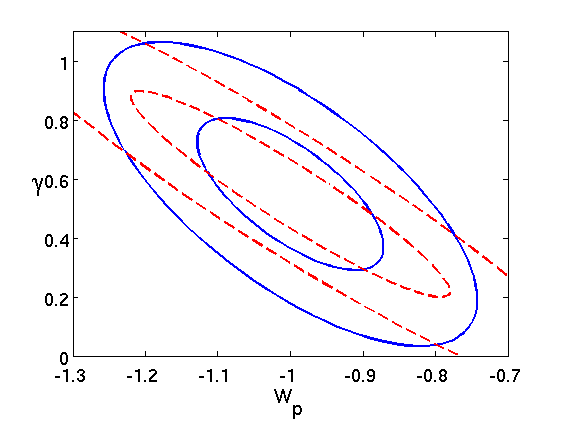}
\includegraphics[width=80mm]{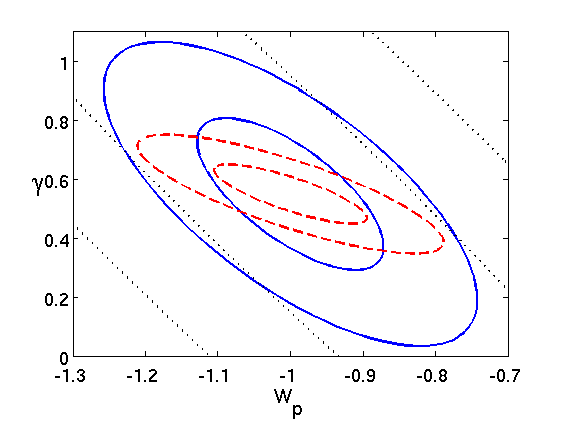}
\caption{ \label{fig:fig1} \textit{Left:} Joint constraints 
on modified gravity and dark energy from a
combination of the Cosmic Microwave Background and Large Scale
Structure. The solid contours represent the 1- and 2-$\sigma$
constraints for a $10 \gpcubed$ redshift survey at $z=1$, with
$\bar{n}=10^{-4}h^{3}\mathrm{Mpc}^{-3}$, and combined with the DETF
Planck Fisher matrix. If we had chosen to neglect the Alcock-Paczynski
effect from the redshift distortions, but leaving the BAO information intact as discussed in the text, we would arrive at the dashed contours. In both cases, a weak prior is applied to $\sigma_p$, on the basis it may be well measured on scales much smaller than those considered here. \textit{Right:} Varying the redshift bin of the survey from $z=0.5, 1, 2$, shown as dashed, solid and dotted respectively.}
\end{figure*}

\begin{figure}
\includegraphics[width=80mm]{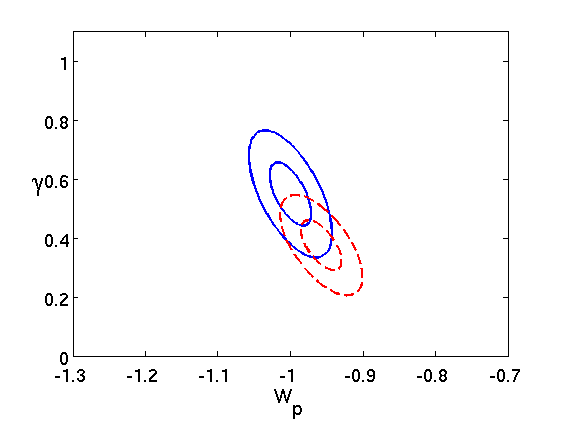}
\caption{\label{fig:curvature} The same survey
specifications as in Fig \ref{fig:fig1}, but now assuming a flat
universe. The dashed contours illustrate the bias induced by an actual
value of $\Omega_k=-0.01$. }
\includegraphics[width=80mm]{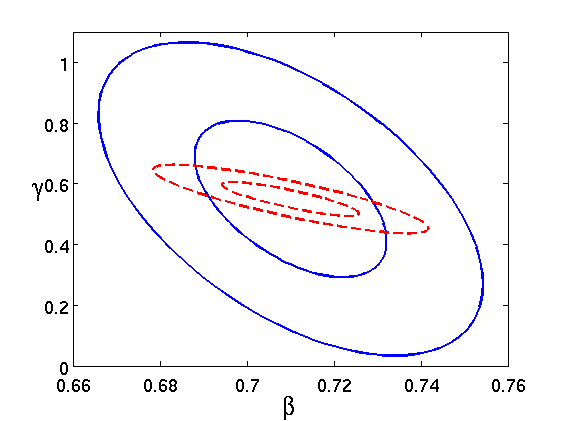} 
\caption{ \label{fig:fig5}  The substantial error in
$\gamma$ that can arise from a relatively small uncertainty in
$\beta$. The dataset is the same as Fig \ref{fig:fig1}, and the
dashed line is for a fixed $w=-1$.}
\end{figure}

\begin{figure}
\includegraphics[width=80mm]{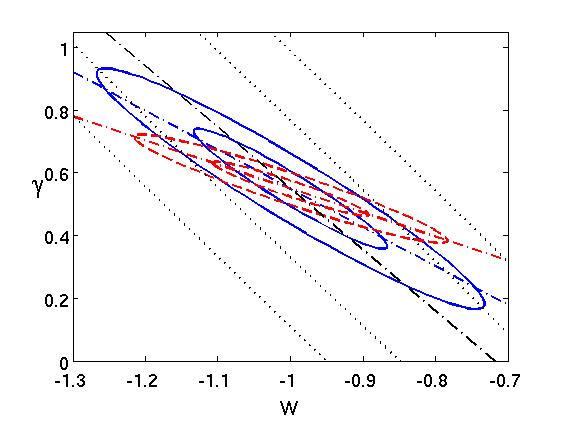}
\caption{ \label{fig:fig7} The dash-dot lines highlight the degeneracy directions for redshift distortions at $z=0.5, 1, 2$, as predicted by (\ref{eq:degen}). The solid contours correspond to the same dataset as in Figure \ref{fig:fig1}, and we have fixed $w_a=0$ to provide a consistent comparison. }
\end{figure}

\section{Results} \label{sec:results}

Once we are in possession of a full Fisher matrix, marginalization
can be performed in the usual analytic manner, in order to isolate
the constraints on the parameters of interest. The resulting
confidence contours are shown in Figures
~\ref{fig:fig1}-\ref{fig:fig7}. Here we illustrate a survey that should be
feasible ($N=10^6$ redshifts). The striking aspect of these plots,
which is the main result of our paper, is that there is a strong
degeneracy between $w$ and $\gamma$, in the sense that less
negative $w$ requires a smaller value of $\gamma$; the slope of
this degeneracy depends on redshift, but is always in this sense.

Huterer \& Linder \cite{2007PhRvD..75b3519H} highlighted the bias
induced in $w$ by neglecting $\gamma$. Here we would stress that
the reverse may also be true, a deviation in general relativity
may be erroneously inferred by neglecting a deviation from $w=-1$.
To state the issue more simply: the conditional errors in
$w$ and $\gamma$ may seriously underpredict the total
error in either parameter when marginalizing over the
unknown value of the other. In the examples we have shown,
this almost doubles the error.

\subsection{Effect of Alcock-Paczynski}

One of the main differences between this and most earlier works is the
inclusion of the full distortion of the power spectrum, as given by
(\ref{eq:fullpk}). This includes both the BAO and RSD information,
although these have previously been discussed as separate effects.
To clarify this, consider ({\ref{eq:fullpk}) again. The first term,
\[
P'_{\rm gal}(k')=\frac{1}{\fpe^{2}\fpa}b{}^2P_{m}\left(\frac{k'}{\fpe}\sqrt{1+\mu'^{2}\left(\frac{1}{F^{2}}-1\right)}\;\right),
\]
accounts for BAO (plus the information in the overall
curvature of the power spectrum, which we will not attempt
to separate out). This appears different, but has the same
content as the standard approach, which is to compute
$P'_{\rm gal}(k')$ and hence the acoustic scale using one geometry only,
but then to argue that the scale for other geometries should
change $\propto D_V\equiv [(1+z)^2 D_A^2 cz/H]^{1/3}$. This approximate
scaling applies only in the absence of RSD, however; since these are
always present whether or not the analysis focuses on BAO only,
the full analysis is to be preferred.

RSD have frequently been discussed in isolation, using the Kaiser
formula. One might have thus been tempted to approach a combined
BAO+RSD analysis by adopting an incorrect model that treats BAO
as above, together with RSD without the Alcock-Paczynski corrections:
\[
\eqalign{
P'_{\rm gal}(k')&=\frac{1}{\fpe^{2}\fpa}b{}^2P_{m}\left(\frac{k'}{\fpe}\sqrt{1+\mu'^{2}\left(\frac{1}{F^{2}}-1\right)}\;\right) \cr
&\times \left[1+\beta \mu'^{2}\right]^{2}D\left(k'_\parallel \sigma_{p}\right).
}
\]
It is instructive to compare this form with the correct power
spectrum, in order to demonstrate the impact of the Alcock-Paczynski corrections.
This modification
results in the dashed contours shown in Figure ~\ref{fig:fig1}, which
illustrates how neglecting the geometric distortion of $P'(k)$ leads
to erroneously small conditional errors on both $w$ and $\gamma$.
Furthermore, the lack of anisotropic amplification reduces our
capability of distinguishing between deviations in $w$ and $\gamma$,
thereby increasing the (negative) covariance between the two
parameters.

In earlier work, both Sapone \& Amendola \cite{2007arXiv0709.2792S} and Stril et al. \cite{2009arXiv0910.1833S} address this issue, although here our analysis extends to include parameters such as $\Omega_k$ and $\sigma_p$, of which we find the former provides a substantial impact.  Wang \cite{2008JCAP...05..021W} explored the potential for galaxy redshift survey  to measure the linear growth $f(z)$, including the Alcock-Paczynski effect. Yet this parameterisation conceals uncertainties in $H(z)$, so here we focus on purely growth-dependent term, $\gamma$.

We note in passing that it is not so straightforward to achieve the
converse of a BAO ``wiggles-only" analysis and cleanly isolate the RSD signal alone. For instance, if we were to null
the isotropic component
\[
\bar{P}(k_{\parallel},k_{\perp}) = \frac{ P(k_{\parallel},k_{\perp})}{\int P(\bf{k}) \ud \mu}  ,
\]
this removes an essential component of the signal we
require, namely the amplitude of the power spectrum itself.

\subsection{Redshift evolution}

As we progress towards higher redshifts $\Omega_m(z)$ approaches
unity, and so for a fixed fractional error on $\beta$ we arrive at a
larger error $\delta \gamma$, as given by

\[
\delta\gamma=\frac{1}{\ln\left(\om(z)\right)} \frac{\delta f}{f}  .
\]

\noindent This effect is illustrated in the right panel of Figure
\ref{fig:fig1}. A tilting in the degeneracy direction is also induced.

\subsection{Curvature}

Non-zero curvature not only contributes to the Alcock-Paczynski
squashing, but invokes further uncertainty in $\Omega_m(z)$, which in
turn enlarges the uncertainty in $\gamma$. Figure
\ref{fig:curvature} demonstrates the dangers associated with assuming
a flat cosmology, where even a modest deviation from flatness
$\Omega_k = 0.01$ can be seen to generate a significant bias in the
estimation of both $w$ and $\gamma$.




\subsection{Errors on $\beta$} \label{sec:fore}

Two fitting functions have recently been proposed to predict the
precision with which $\beta$ may be measured. Guzzo et
al. \cite{2008Natur.451..541G} utilised the correlation function in
real-space, and found the error on $\beta$ was well described by
\[
\frac{\delta\beta}{\beta}=\frac{50}{V^{0.5}\bar{n}^{0.44}}  ,
\]
while White et al. \cite{2009MNRAS.tmp..925W} considered the analysis
in Fourier space, noting
\[
\frac{\delta\beta}{\beta}=b^{-1}\left[\beta^{2}F_{bb}^{-1}-2\beta F_{bf}^{-1}+F_{ff}^{-1}\right]^{1/2}  ,
\]
which exhibits the same scaling with volume, but is rather more
pessimistic at high number densities.

When fixing the background cosmology, our findings are consistent with
White et al.

\subsection{Degeneracy direction}

To establish the expected direction of degeneracy in the $w - \gamma$
plane, we consider the partial derivatives of the relevant parameters
\[ \label{eq:degen}
\frac{\partial \gamma}{\partial w} = - \frac{\displaystyle \frac{\partial \ln f^{\phantom{A}}}{\partial w} +
\frac{\partial \ln g}{\partial w}}{\displaystyle \frac{\partial \ln f}
{ \partial \gamma_{\phantom{\gamma}}} + \frac{\partial \ln g}{\partial \gamma}} ,
\]
where $g \equiv \sigma_8(z) / \sigma_8 (z_{\rm CMB})$. This predicted
degeneracy gradients are plotted as dotted lines in Figure ~\ref{fig:fig7}, and can be seen to closely align with
the redshift distortion contours for a constant equation of state. The dataset matches that of Figure
~\ref{fig:fig1}.

A simple qualitative interpretation of the slope direction is that a more positive value of $w(z)$ generates a lower $\Omega_m(z)$, which in turn requires a lower value of $\gamma$ in order to maintain the same value of $f$.

\section{Figure of Merit} \label{sec:merit}

In recent work by the Joint Dark Energy Mission Figure of
Merit Science Working Group \cite{2009arXiv0901.0721A}, the
relative merits of future surveys are quantified by separately
considering the errors on the dark energy equation of state (in
the form of eigenmodes) and the modified growth index
$\Delta\gamma$. However as we have seen, these quantities can clearly exhibit
significant covariance. To compensate for this, a simple
prescription could be adopted in terms of the marginalised Fisher elements, analogous to that previously used for $w_0$ and
$w_a$,

\[ \label{eq:fom}
{\rm FoM}= \sqrt{F_{ww}F_{\gamma\gamma}-F_{w\gamma}^{2}} .
\]

\noindent Note we have omitted the factor of approximately $6 \pi$ which would  reduce this to the area within the $95\%$ confidence contours. Some examples of this FoM are presented in Table \ref{tab:1}, for a selection of survey volumes and redshifts.

We also stress that the quantity $\gamma$ is just as likely to exhibit
redshift variation as $w$, which raises the question: \emph{where}
are we measuring its value?  The functional form is such that $z<2$ is
strongly preferred, since at higher redshifts $\Omega_{m}\sim 1$ and
$\gamma$ is unable to exert much influence. This reflects our prior
that regions of low $\Omega_\Lambda$ are less likely to demonstrate
unusual activity in the growth rate. Dark energy is only known to
exist as a low-redshift phenomenon, and as such unearthing the growth
rate in this era presents a most enticing prospect. This issue is
elaborated in a companion paper \cite{simp09}.


\begin{table}
\begin{center}
\begin{tabular}{c>{\bfseries}c|c|c|c|c|}
\cline{3-6}
& & \multicolumn{4}{|c|}{Mean Redshift} \\ \cline{3-6}
  & & 0.5 & 1 & 1.5 & 2 \\ \cline{1-6}
\multicolumn{1}{|c|}{}                        &
\multicolumn{1}{|c|}{1}   & 16.8 & 3.8 & 0.6 & 0.3      \\ \cline{2-6}
\multicolumn{1}{|c|}{Volume}                        &
\multicolumn{1}{|c|}{10} & 158.0 & 35.4 & 5.8 & 2.6      \\ \cline{2-6}
\multicolumn{1}{|c|}{$(\gpcubed)$}                        &
\multicolumn{1}{|c|}{100} &  - & 329.5 & 55.7 & 25.2   \\ \cline{1-6}
\end{tabular}
\end{center}
\caption{\label{tab:1}The Figure of Merit, as defined by (\ref{eq:fom}), for various permutations of volume and redshift, for the case of a single redshift bin and with a number density $\bar{n} = 10^{-3}h^{3}\mathrm{Mpc}^{-4}$ \$.}
\end{table}

\begin{table}
\begin{center}
\begin{tabular}{c>{\bfseries}c|c|c|c|c|}
\cline{3-6}
& & \multicolumn{4}{|c|}{Mean Redshift} \\ \cline{3-6}
  & & 0.5 & 1 & 1.5 & 2 \\ \cline{1-6}
\multicolumn{1}{|c|}{}                        &
\multicolumn{1}{|c|}{1}   & 175.8 & 62.4 & 25.3 & 12.0       \\ \cline{2-6}
\multicolumn{1}{|c|}{Volume}                        &
\multicolumn{1}{|c|}{10} & 1122.3 & 405.2 & 166.6 & 80.8      \\ \cline{2-6}
\multicolumn{1}{|c|}{$(\gpcubed)$}                        &
\multicolumn{1}{|c|}{100} &  - & 1514.1 & 606.9 & 311.2   \\ \cline{1-6}
\end{tabular}
\end{center}
\caption{\label{tab:2} The same Figure of Merit as in Table \ref{tab:1}, but now under the assumption of a flat universe.}
\end{table}


%

\section{Conclusions}

By relaxing the common assumption of a fixed background cosmology, we
have highlighted some of the difficulties encountered when attempting
to study gravity via the bulk motion of galaxies. Rather than a pure
probe of structure, redshift distortions also comprise a geometric
component. This enters at the stage of converting the true
observables, angles and redshifts, into distances and Fourier
modes. Furthermore, when determining the growth index $\gamma$ it is
essential that its corresponding radix $\Omega_m(z)$ is well
determined. With these two factors in mind, it appears unlikely that
the galaxy power spectrum alone could provide conclusive evidence against
General Relativity.

To converge on the true underlying cosmology, iterating over a value
for $\Omega_m$ has proved adequate for current data. However with the
greater degrees of freedom required to test relativity ($w_0, w_a,
\Omega_k$), the available volume of parameter space appears too
great. Fortunately future data will inevitably be accompanied by
improved measurements of the baryon acoustic oscillations. Ironically
the squashing effect that empowers the BAO is the very same
Alcock-Paczynski effect that confounds the redshift distortions.

One concern in the formalism may be the assumption of
scale-independence for both the growth and bias. More physically motivated
forms of modified gravity, such as $f(R)$ models,
\cite{2007PhRvD..76j4043H}, lead to rather different scale-dependent
growth factors. However, as highlighted in \cite{2007PhRvD..76j4043H},
such models also generate very prominent deviations on intermediate
scales, which would become more immediately apparent.

Nevertheless, neglect of these issues is more likely to lead to a bias
in the results of analyses that assume scale-independent effects,
rather than changing their statistical precision. In this work,
we have concentrated on the latter aspect, and our main conclusion
is that the parameters $\gamma$ and $w_p$ will generally be strongly
anti-correlated. We therefore suggest that a natural Figure of Merit
for future experiments in fundamental cosmology should be the reciprocal
of the area of the error contour in the $\gamma - w_p$ plane.

\noindent{\bf Acknowledgements} \\
We thank Luigi Guzzo for many helpful comments on an earlier
draft of this paper, and also Thomas Kitching and Will Percival for several productive discussions. 
FS was supported by an STFC Rolling Grant.

\bibliography{M:/Routines/dis}

\end{document}